# Extracting Non-Gaussian Governing Laws from Data on Mean Exit Time

Yanxia Zhang[a,&], Jinqiao Duan[b], Yanfei Jin[a,*], Yang Li[c,&]

[a] *Department of Mechanics, Beijing Institute of Technology, Beijing 100081, China*

[b] *Department of Applied Mathematics, College of Computing, Illinois Institute of Technology, Chicago, Illinois 60616, USA*

[c] *State Key Laboratory of Mechanics and Control of Mechanical Structures, College of Aerospace Engineering, Nanjing University of Aeronautics and Astronautics, Nanjing 210016, China*

[&] *Visiting PhD. Student at Department of Applied Mathematics, College of Computing, Illinois Institute of Technology, Chicago, Illinois 60616, USA*

**ABSTRACT:** Motivated by the existing difficulties in establishing mathematical models and in observing the system state time series for some complex systems, especially for those driven by non-Gaussian Lévy motion, we devise a method for extracting non-Gaussian governing laws with observations only on mean exit time. It is feasible to observe mean exit time for certain complex systems. With the observations, a sparse regression technique in the least squares sense is utilized to obtain the approximated function expression of mean exit time. Then, we learn the generator and further identify the stochastic differential equations through solving an inverse problem for a nonlocal partial differential equation and minimizing an error objective function. Finally, we verify the efficacy of the proposed method by three examples with the aid of the simulated data from the original systems. Results show that the method can apply to not only the stochastic dynamical systems driven by Gaussian Brownian motion but also those driven by non-Gaussian Lévy motion, including those systems with complex rational drift.

**Stochastic differential equations arise in modeling dynamical systems under random fluctuations in many fields. However, it is sometimes challenging to establish the mathematical models or governing laws, due to lack of scientific understanding. Fortunately, advances in data science and machine learning have led to a few data-driven approaches to extract these models with the aid of massive datasets. Currently, most of these approaches are limited to data of system state time series and mainly focus on systems driven by Gaussian Brownian motion. However, for some complex dynamical systems, under heavy-tailed, intermittent, non-**

[*] Corresponding author.
  *E-mail address:* 3120170019@bit.edu.cn (Y. Zhang), duan@iit.edu (J. Duan), jinyf@bit.edu.cn (Y. Jin), li_yang@nuaa.edu.cn (Y. Li).

**Gaussian fluctuations, it is sometimes too costly or difficult to observe the system state time series. Besides, mean exit time may be observed or measured feasibly in some systems. Therefore, in order to deal with the existing difficulties in establishing mathematical models and in observing the system state time series, it is of great interest to identify stochastic differential equations with observations only on mean exit time.**

## I. INTRODUCTION

Stochastic differential equations (SDEs) arise in modeling dynamical systems under random fluctuations in many fields, including biophysics, chemistry, mechanical and electrical engineering, and environment science[1-3]. In general, mathematical models of systems based on the fundamental governing laws help us uncover the complicated nonlinear dynamics. However, for some complex systems, e.g., the systems driven by non-Gaussian Lévy motion, it is often difficult and challenging to build mathematical models or governing law because of the lack of scientific understanding. Fortunately, advances in data science and machine learning have led to new progresses in understanding the complex dynamical systems with the aid of massive datasets. Recently, a few data-driven approaches to identify SDEs from data have been proposed, including the Sparse Identification of Nonlinear Dynamics (SINDy)[4-6], data-driven approximation of the Koopman generator[7-11] and parameters estimation[12-14]. The SINDy approach can be regarded as a milestone for data-driven discovery of dynamical systems. It combines ideas from sparse regression and compressed sensing to discover the terms of differential equations[4], and then is extended to identify parameters of a stochastic system by Kramers–Moyal formulae[5]. The system identification approach via data-driven approximation of the Koopman generator[7] is carried out with the aid of extended dynamic mode decomposition. However, most of these approaches are limited to data from the observations of system state time series or focus on systems driven by Gaussian Brownian motion.

For some complex stochastic dynamical systems, especially those driven by non-Gaussian Lévy motion, it is sometimes too costly or difficult to observe the system state time series over a long period of time[15], but it is feasible to observe other system quantity such as mean exit time. Lévy motion is a non-Gaussian process with heavy-tailed distribution and has been widely used in modeling various systems with heavy-tailed fluctuations[16-18]. Mean exit time, also called as mean residence time or mean first passage time, is a significant deterministic quantity and has been observed or measured in many systems such as fluid, industrial, chemical and physiological systems[19-26]. For instance, the mean residence time of a fluid particle can be measured by "pulse experiment[22]", which consists of introducing a pulse of tracer in the inlet to the domain, while measuring continuously the tracer concentration at a given point. The mean residence time of a fluid in water storage tanks and wastewater lagoons is often measured by Lagrangian simulation of particles injected into the domain[23]. Novotny et al.[24] determined the mean residence time of Xenon in intact and surgically isolated muscles by recording the time-dependent background radioactivity with a detector positioned midway between the two detectors over the tissues. Nasserzadeh et al.[25] have clearly demonstrated that residence times of gas particles in large incinerator plants can be measured by a pseudo-random binary signal tracer technique successfully and economically. Moreover, Ghirelli and Leckner[26] proposed a transport equation for the local residence time of a fluid applying to transient flows and the equation can be used to measure the residence time of a generic reacting chemical species. Therefore, it is feasible to measure or observe the mean exit time in many systems. With observations on mean exit time, Gao and Duan[27] have successfully quantified the model parameters in dynamical systems driven by non-Gaussian Lévy stable noise. Therefore, it is desirable to identify stochastic differential equations with observations on mean exit time.

The purpose of this paper is to extract governing laws with observations only on mean exit time for systems driven by either Gaussian Brownian motion or non-Gaussian Lévy motion. The highlight of this study is that an inverse problem of nonlocal partial differential equation is solved via minimization with observations only on mean exit time. The proposed method in this paper is beneficial to systems for which mean exit time is more feasible to observe. The remainder of this paper is organized as follows. In Sec. II, we present the theoretical background of the generator of system process driven by Brownian motion and Lévy motion, and the mean exit time. In Sec. III, we devise a framework to identify stochastic differential equations with the observations only on mean exit time. Then, we present three examples to verify our method in Sec. IV and conclude with discussion in Sec. V.

## II. PRELIMINARIES

### A. Lévy motion

We consider a stochastic dynamical system under Brownian motion and Lévy motion modeled by a scalar SDE

$$d X_t = f(X_t) d t + \sqrt{\sigma} d B_t + d L_t^\alpha, \quad X_0 = x, \tag{1}$$

where $f(X_t)$ is the drift (or vector field) of system state process $X_t$, variance $\sigma$ is the positive diffusion coefficient of Gaussian Brownian motion $B_t$, and $L_t^\alpha$ is a non-Gaussian α-stable symmetric Lévy motion with generating triplet $(0, 0, \varepsilon v_\alpha)$. The jump diffusion coefficient $\varepsilon \geq 0$, and it is Gaussian case when $\varepsilon = 0$. The jump measure of the Lévy motion is

$$v_\alpha(d y) = C_\alpha |y|^{-(1+\alpha)} d y, \tag{2}$$

where $\alpha \in (0,2)$ denotes the power parameter or stability index, $C_\alpha = \frac{\alpha}{2^{1-\alpha}\sqrt{\pi}} \frac{\Gamma(1/2+\alpha/2)}{\Gamma(1-\alpha/2)}$. For more details see Refs. 1 and 28.

The generator for the solution process $X_t$ is

$$Ag = f(x)g'(x) + \frac{\sigma}{2}g''(x) + \varepsilon \int_{\mathbb{R}\setminus\{0\}} [g(x+y) - g(x) - I_{\{|y|<1\}} y g'(x)] \nu_\alpha(dy), \quad (3)$$

where $I_S$ denotes the indicator function of the set $S$, that is

$$I_S(y) = \begin{cases} 1, & y \in S; \\ 0, & y \notin S. \end{cases} \quad (4)$$

Due to the rotational invariance of the symmetric α-stable Lévy process with triplet $(0, 0, \varepsilon \nu_\alpha)$, the generator can be reformed as

$$Ag = f(x)g'(x) + \frac{\sigma}{2}g''(x) + \varepsilon \int_{\mathbb{R}\setminus\{0\}} [g(x+y) - g(x)] \nu_\alpha(dy). \quad (5)$$

**B. Mean exit time**

For a bounded domain $D \in R$ (with boundary $\partial D$), the first exit time $\tau_D$ for the solution orbit starting at $x \in D$ is defined as

$$\tau_D(\omega) \triangleq \inf\{t > 0 : X_0 = x, X_t \in \partial D\}. \quad (6)$$

Then the mean exit time of a particle initially at $x \in D$ escaping from $D$ is defined as

$$u(x) \triangleq E\tau_D(\omega). \quad (7)$$

It turns out that the mean exit time, under the uniform ellipticity condition[1], satisfies the following elliptic nonlocal partial differential equation

$$\begin{aligned} Au(x) &= -1, \\ u\big|_{D^c} &= 0, \end{aligned} \quad (8)$$

where $A$ is the generator defined in Eq. (3), and $D^c$ is the complement of the domain $D$.

**III. METHOD**

In order to identify stochastic differential equations with observations only on mean exit time, we devise a method through an inverse problem and minimization in this section. We first calculate the regression function of mean exit time with given observations by utilizing a sparse regression technique in the least squares sense, and then derive the drift expression by solving an inverse problem of nonlocal partial differential equation. A minimization scheme of the constructed error objective function is then used to learn the noise coefficients and the drift for stochastic dynamical equations in subsec. C. Finally, we outline the proposed algorithm for identifying systems from data of mean exit time in subsec. D.

**A. Sparse regression with observations on MET**

Assume that we have observed $n$ data points of MET, denoted by $\mathbf{U}_{ob}$, in $D=(a,b)$. In order to satisfy the boundary condition in Eq. (8), we construct the basis function as

$$\varphi(x) = [\varphi_1(x),\cdots,\varphi_K(x)] = (x-a)(b-x)[1, x, x^2,\cdots, x^{K-1}]. \tag{9}$$

Then, the MET can be approximated as

$$u(\mathbf{x}) \approx \sum_{k=1}^{K} c_k \varphi_k(\mathbf{x}). \tag{10}$$

Consequently, we have

$$B\mathbf{c} = \mathbf{U}_{ob} \tag{11}$$

with

$$B = \begin{bmatrix} \varphi_1(x_1) & \cdots & \varphi_K(x_1) \\ \vdots & \ddots & \vdots \\ \varphi_1(x_n) & \cdots & \varphi_K(x_n) \end{bmatrix}. \tag{12}$$

In general, Eq. (11) might have no solutions as there are more expressed equations than variables. Thus, it needs to be solved in the least squares sense, i.e., $\arg\min_{\mathbf{c}} \|B\mathbf{c} - U_{ob}\|_2$. The solution $\mathbf{c}$ of Eq. (11) can be leaded as

$$\mathbf{c} = \left(B^T B\right)^{-1} \left(B^T \mathbf{U}_{ob}\right). \tag{13}$$

It should be noted that here $\mathbf{c}$ is a non-sparse solution. For the goal of finding the minimal functional form of MET without loss of reliability representing the data among the large number of possibilities provided in the function dictionary, we need to enforce the sparsity of $\mathbf{c}$ by the iterative thresholding algorithm[29]. That is, given a pre-defined threshold parameter $\lambda$ which needs to be adjusted appropriately, set the coefficients of $\mathbf{c}$ less than value $\lambda$ to zero and regress the remaining coefficients. This process is iterated until no coefficient less than $\lambda$. The sparsity result of $\mathbf{c}$ is denoted by $\tilde{\mathbf{c}}$. Thereby, we can obtain the approximated MET as a function of $x$, i.e.,

$$u_f(x) = \sum_{l=1}^{L} \tilde{c}_l \varphi_l(x), \quad x \in (a,b). \tag{14}$$

**B. Inverse problem**

In order to get the drift, we first need to solve the inverse problem of the nonlocal partial differential equation (8). For simplify, $u_f(x)$ can be furthermore reformed as

$$u_f(x) = \sum_{l=1}^{L} \tilde{c}_l \varphi_l(x) = \sum_{j=0}^{N} r_j x^j, \quad x \in (a,b). \tag{15}$$

where $r_j$ ($j = 1, 2, \cdots N$) can be calculated by $\tilde{\mathbf{c}}$, and $u_f(x) = 0$ for $x \notin (a,b)$.

Based on Eq. (8), we have

$$Au_f = f(x)u_f'(x) + \frac{\sigma}{2}u_f''(x) + \varepsilon C_\alpha \int_{\mathbb{R}\setminus\{0\}} \frac{u_f(x+y) - u_f(x)}{|y|^{1+\alpha}} dy = -1. \tag{16}$$

Due to $u_f(x) = 0$ for $x \notin (a,b)$, Eq. (16) can be solved by decomposing the integral into four parts, i.e. $\int_R = \int_{-\infty}^{a-x} + \int_{a-x}^{x-a} + \int_{x-a}^{b-x} + \int_{b-x}^{\infty}$. Then, we have

$$\begin{aligned} &f(x)u_f'(x) + \frac{\sigma}{2}u_f''(x) - \frac{\varepsilon C_\alpha}{\alpha}\left[\frac{1}{(x-a)^\alpha} + \frac{1}{(b-x)^\alpha}\right]u_f(x) \\ &+ \varepsilon C_\alpha \left[\int_{a-x}^{x-a} \frac{u_f(x+y) - u_f(x)}{|y|^{1+\alpha}} dy + \int_{x-a}^{b-x} \frac{u_f(x+y) - u_f(x)}{|y|^{1+\alpha}} dy \right] = -1. \end{aligned} \tag{17}$$

For convenience, set $\int_{a-x}^{x-a} = M_1$ and $\int_{x-a}^{b-x} = M_2$ in short. Substituting the approximated function $u_f(x)$ of Eq. (15) into these two integral terms, respectively, we can get

$$M_1 = \sum_{j=0}^{N} r_j \int_{a-x}^{x-a} \frac{(x+y)^j - x^j}{|y|^{1+\alpha}} dy = 2\sum_{j=2}^{N}\sum_{k=2m}^{j} \frac{C_j^k r_j x^{j-k}}{k-\alpha}(x-a)^{k-\alpha}, \tag{18}$$

$$M_2 = \sum_{j=0}^{N} r_j \int_{x-a}^{b-x} \frac{(x+y)^j - x^j}{|y|^{1+\alpha}} dy = \sum_{j=1}^{N}\sum_{k=1}^{j} \frac{C_j^k r_j x^{j-k}}{k-\alpha}\left[(b-x)^{k-\alpha} - (x-a)^{k-\alpha}\right], \tag{19}$$

where $C_{l-1}^k = \frac{(l-1)!}{k!(l-1-k)!}$, $C_0^0 = 1$, and $m \in N^+$.

Noting here that when $\alpha = 1$, the term $M_2$ should be calculated as

$$M_2 = \sum_{j=1}^{N}\sum_{k=2}^{j} \frac{C_j^k r_j x^{j-k}}{k-\alpha}\left[(b-x)^{k-\alpha} - (x-a)^{k-\alpha}\right] + \sum_{j=1}^{N} jr_j x^{j-1} \ln\left|\frac{b-x}{x-a}\right|. \tag{20}$$

Consequently, we can express the drift term as a function of ($x, \sigma, \alpha$) by substituting Eqs. (15,18-20) into Eq. (17) as

$$f(x,\sigma,\alpha) = u_f'(x)^{-1}\left\{-\frac{\sigma}{2}u_f''(x) + \frac{\varepsilon C_\alpha}{\alpha}\left[\frac{1}{(x-a)^\alpha} + \frac{1}{(b-x)^\alpha}\right]u_f(x) - \varepsilon C_\alpha(M_1 + M_2) - 1\right\}. \tag{21}$$

Noting that $f \to \infty$ when $u_f'(x) \to 0$. Thus, when solving it by discrete form in the following work, we apply the L'Hôpital's rule here for $|u_f'(x)| < \theta$ ($0 < \theta \ll 1$), i.e.,

$$f(x,\sigma,\alpha) = u_f''(x)^{-1} \left\{ -\frac{\sigma}{2} u_f''(x) + \frac{\varepsilon C_\alpha}{\alpha} \left[ \frac{1}{(x-a)^\alpha} + \frac{1}{(b-x)^\alpha} \right] u_f(x) - \varepsilon C_\alpha (M_1 + M_2) - 1 \right\}'. \quad (22)$$

**C. System identification**

Because of the unknown parameters $\sigma$ and $\alpha$ in the learned drift term in Eqs. (21,22), we construct an error objective function as

$$G(\sigma,\alpha) = \frac{\|U_L(x,\sigma,\alpha) - \mathbf{U}_{ob}\|_2^2}{\|\mathbf{U}_{ob}\|_2^2} \quad (23)$$

where $\mathbf{U}_{ob}$ is the observed discrete data on MET, $U_L(x,\sigma,\alpha)$ is the learned MET obtained by substituting Eq. (21) into Eq. (8). Due to the complexity of the generator $A$ especially at the present of Lévy motion, the learned MET $U_L$ needs to be solved by discrete form with the finite difference scheme proposed by Gao et al. in Ref. 30. For any $\sigma > 0$ and $\alpha \in (0,2)$, we have the following discretized equation for MET from $Au(x) = -1$

$$C_h \frac{U_{j-1} - 2U_j + U_{j+1}}{h^2} + \tilde{f}(x_j,\sigma,\alpha) \frac{U_{j+1} - U_{j-1}}{2h} - \frac{\varepsilon C_\alpha U_j}{\alpha} \left[ \frac{1}{(x_j-a)^\alpha} + \frac{1}{(b-x_j)^\alpha} \right] + \varepsilon C_\alpha h \sum_{\substack{k=-J-j \\ k \neq 0}}^{J-j} {}'' \frac{U_{j+k} - U_j}{|x_k|^{1+\alpha}} = -1, \quad (24)$$

where $C_h = \frac{\sigma}{2} - \varepsilon C_\alpha \zeta(\alpha - 1) h^{2-\alpha}$ with the Riemann zeta function $\zeta$, $j = -J+1, \cdots, -1, 0, 1, \cdots, J-1$ and $h = (b-a)/2J$.

For guaranteeing the smoothness of drift term for any given values of $\sigma$ and $\alpha$, we first fit curve to the discrete points from Eq. (22) by polynomial basis function $\psi(x) = [\psi_1(x), \cdots, \psi_K(x)] = [1, x, x^2, \cdots, x^{K-1}]$ and take the approximated drift $\tilde{f}(x_j,\sigma,\alpha)$ as

$$\tilde{f}(x_j,\sigma,\alpha) \approx \sum_{k=1}^{K} \tilde{q}_k \psi_k(x_j), \quad j = 1, 2, \ldots, n, \quad (25)$$

in which, $\tilde{\mathbf{q}}$ is the sparsity result of $\mathbf{q}$ calculated as

$$\begin{aligned} \mathbf{q} &= (Q^T Q)^{-1} (Q^T f(\mathbf{x},\sigma,\alpha)), \\ Q &= \begin{bmatrix} \psi_1(x_1) & \cdots & \psi_K(x_1) \\ \vdots & \ddots & \vdots \\ \psi_1(x_n) & \cdots & \psi_K(x_n) \end{bmatrix}. \end{aligned} \quad (26)$$

The identification of the learned $\sigma_L$ and $\alpha_L$ can be achieved by minimizing the error objective function, i.e.,

$$[\sigma_L, \alpha_L] = \min_{\sigma \geq 0, \alpha \in (0,2)} G(\sigma,\alpha). \quad (27)$$

Accordingly, the learned drift $\tilde{f}(x,\sigma_L,\alpha_L)$ can be obtained by the learned $\sigma_L$ and $\alpha_L$. Hereto, we have completed the learning of generator $A$. Thus, we obtain the data-driven stochastic differential equations.

**D. Algorithm**

The algorithm is implemented in Matlab and its basic procedure is outlined in Algorithm 1.

---

**Algorithm 1** Identifying SDEs with observations on mean exit time

**Require**: Basis functions $\varphi(x)$ and $\psi(x)$.

**Input**: A data set of observations on mean exit time $\mathbf{U}_{ob}$ in $D = (a,b)$.

1. Calculate the function expression $u_f(x) = \sum_{l=1}^{L} \tilde{c}_l \varphi_l(x) = \sum_{j=0}^{N} r_j x^j$ by Eqs. (9)-(14).
2. **for** any $\sigma \in (0,2)$ and $\alpha \in (0,2)$
3. Calculate the drift $f(x,\sigma,\alpha)$ by Eq. (21)
4. Construct the polynomial basis function and obtain the approximated drift $\tilde{f}(x,\sigma,\alpha)$ by Eq. (25)
5. Calculate the error objective function $G(\sigma,\alpha)$ by Eq. (23) for all combinations of $(\sigma,\alpha,\tilde{f}(x,\sigma,\alpha))$
6. **end**
7. find($G$==min($G$))
8. System identification by Eqs. (27) and (25).

**Output**: The error objective function and mean exit time.

---

## IV. NUMERICAL EXPERIMENTS

In this section, we present three examples to verify our method with the simulated data from the original systems. We first consider a tri-stable dynamical system driven by Brownian motion, and then a genetic regulatory system with complex rational drift function driven by Brownian motion. Furthermore, we consider a bi-stable dynamical system driven by Lévy motion.

**Example 1.** Consider a tri-stable dynamical system driven by Brownian motion

$$\mathrm{d}X_t = (X_t - 4X_t^3 + 3.5X_t^5)\,\mathrm{d}t + \mathrm{d}B_t. \tag{28}$$

According to Eq. (5) with $\varepsilon = 0$, the generator of this system with drift $f(x) = x - 4x^3 + 3.5x^5$ and diffusion $\sigma = 1$ is

$$Ag = (x - 4x^3 + 3.5x^5)g'(x) + \frac{1}{2}g''(x). \tag{29}$$

Firstly, we take the simulated data from $Au = -1$ as the observations on MET $\mathbf{U}_{ob}$ by difference scheme provided by Eq. (24). The data set has $2J$ points on escaping interval $D = (-1,1)$, where $J = 120$. During the regression procedure, the basis function with order up to 16 is chosen and the threshold parameter is set as $\lambda = 0.01$.

Through the regression procedure of subsection 3.1, the sparse coefficients $\tilde{\mathbf{c}}$ can be got as

$$(\tilde{c}_1, \tilde{c}_3, \tilde{c}_5, \tilde{c}_7, \tilde{c}_9, \tilde{c}_{11}, \tilde{c}_{13}) = (0.9526, -0.0474, 0.2855, -0.3304, 0.2058, -0.2022, 0.0729), others \quad \tilde{c}_l = 0. \tag{30}$$

Furthermore, we can get the approximated function of MET $u_f(x)$ by Eq. (15)

$$u_f(x) = \sum_{j=0}^{N} r_j x^j, \tag{31}$$

where $N = 14$ and

$$(r_0, r_2, r_4, r_6, r_8, r_{10}, r_{12}, r_{14}) = (0.9526, -1.0, 0.3329, -0.6159, 0.5361, -0.408, 0.2751, -0.07288), \text{others} \quad r_j = 0. \tag{32}$$

Consequently, the drift term can be expressed as a function of $(x, \sigma)$ by Eq. (21)

$$f(x,\sigma) = \left[ -\frac{\sigma}{2} \sum_{j=2}^{N} j(j-1) r_j x^{j-2} - 1 \right] \bigg/ \sum_{j=1}^{N} j r_j x^{j-1}. \tag{33}$$

The polynomial basis function with order up to 6 is chosen to calculate the approximated drift $\tilde{f}(x,\sigma)$ for any $\sigma$. Then the learned $\sigma_L$, the optimization value of $\sigma$ minimizing the error objective function shown in Fig. 1(a), can be obtained as

$$\sigma_L = \min_{\sigma \geq 0} G(\sigma) = 1.00. \tag{34}$$

Accordingly, we can obtain the sparse coefficient $\tilde{\mathbf{q}}$ of the learned drift $\tilde{f}(x)$ with the learned $\sigma_L$, as shown in table 1. We can see that the learning result is very close to the true value. Comparation between the mean exit time from the learned SDE and the observations is also present in Fig. 1(b), which can be seen that they are very well in agreement.

Table 1: Identified coefficients for the drift term

| Basis | True | Learning |
| --- | --- | --- |
| 1 | 0 | 0 |
| x | 1 | 1.0002 |
| $x^2$ | 0 | 0 |
| $x^3$ | -4 | -3.9999 |
| $x^4$ | 0 | 0 |
| $x^5$ | 3.5 | 3.4983 |
| $x^6$ | 0 | 0 |

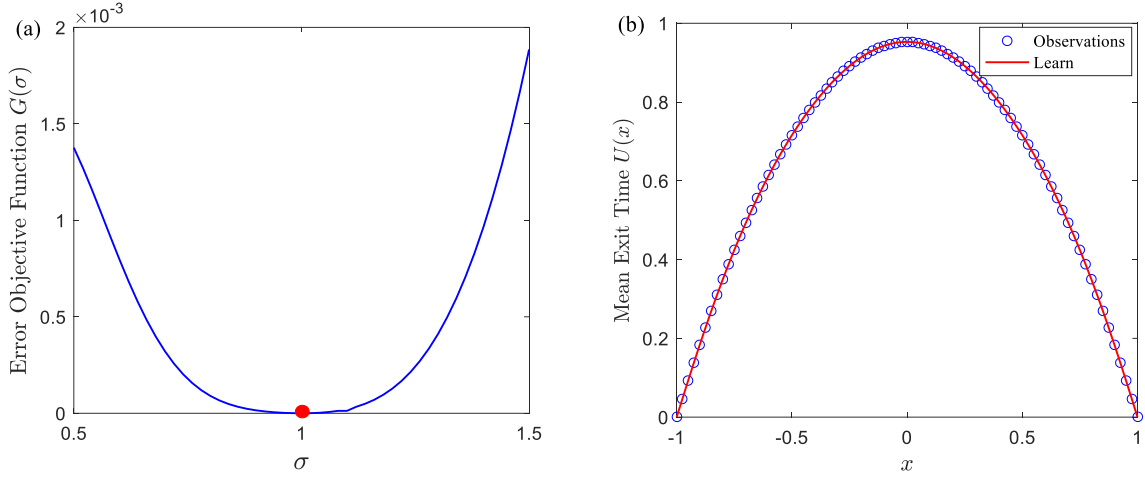

**FIG. 1.** A tri-stable dynamical system with Brownian motion. (a) The error objective function versus $\sigma$. (b) Comparation between the mean exit time from the learned SDE and the observations.

**Example 2.** Consider a genetic regulatory system with complex rational drift function driven by Brownian motion

$$dX_t = (\frac{k_f X_t^2}{X_t^2 + K_d} - k_d X_t + R_{bas})dt + \sqrt{0.5}\,dB_t, \tag{35}$$

where system parameters are $k_f = 6\,\text{min}^{-1}$, $K_d = 10$, $k_d = 1\,\text{min}^{-1}$, $R_{bas} = 0.4\,\text{min}^{-1}$, as in Ref. 31.

According to Eq. (5) with $\varepsilon = 0$, the generator of this system with drift $f(x) = \frac{k_f X_t^2}{X_t^2 + K_d} - k_d X_t + R_{bas}$ and diffusion $\sigma = 0.5$ is

$$Ag = (\frac{k_f X_t^2}{X_t^2 + K_d} - k_d X_t + R_{bas})g'(x) + \frac{1}{4}g''(x). \tag{36}$$

Under the same setting with Example 1, we obtain the observations on MET $\mathbf{U}_{ob}$ from the simulated data set by $Au = -1$.

Similar to Example 1, choosing the polynomial basis function as the dictionary to approximate the drift term, along with the minimization of the error objective function, we can obtain the learned diffusion $\sigma_L \approx 0.47$ and the learned drift

$$\tilde{f}(x) = 0.3552 - 0.9154x + 0.5992x^2 - 0.1007x^3 - 0.0046x^4 + 0.0014x^5. \tag{37}$$

Figure 2(a) shows the variation of the error objective function $G(\sigma)$ with $\sigma$. Figure 2(b) and 2(c) show the comparations between the learned results and the true or the observations in drift term and mean exit time. Results show that the learned results and the true results are very well in agreement. These indicate that our method can also be used for identifying the stochastic dynamical systems with complex rational drift and further characterize the dynamical features with a high accuracy.

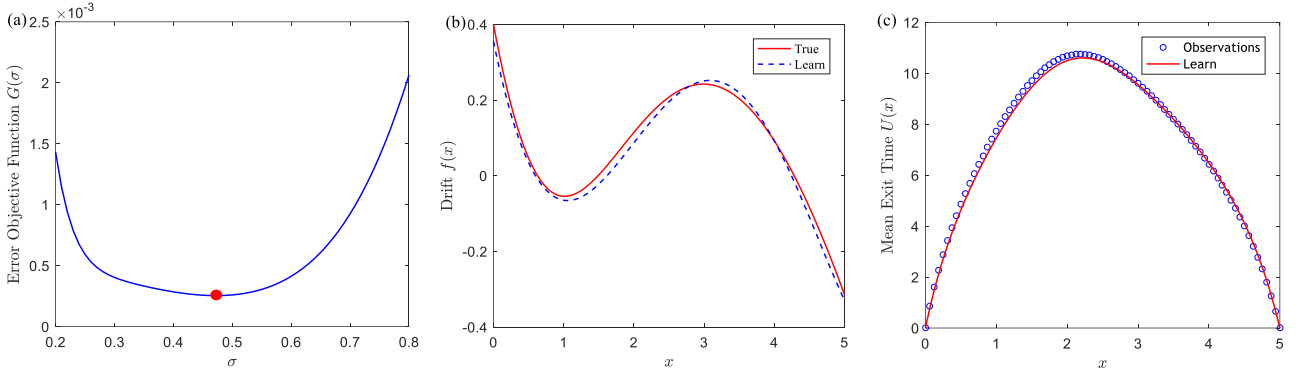

**FIG. 2**. A genetic regulatory system with Brownian motion. (a) The error objective function versus $\sigma$. (b) Comparation between the learned and the true result. (c) Comparation between the mean exit time from the learned SDE and the observations.

**Example 3.** Consider a bi-stable dynamical system driven by Lévy motion

$$dX_t = (X_t - 5X_t^3)dt + \sqrt{0.5}\, dB_t + dL_t^\alpha, \tag{38}$$

where $L_t^\alpha$ is a α-stable symmetric Lévy motion with generating triplet $(0,0,\nu_\alpha)$, and the power parameter $\alpha = 0.6$. Accordingly, with the addition of the drift $f(x) = x - 5x^3$ and diffusion $\sigma = 0.5$, the generator of this system is

$$Ag = (x - 5x^3)g'(x) + \frac{1}{4}g''(x) + C_\alpha \int_{\mathbb{R}\setminus\{0\}} [g(x+y) - g(x)]|y|^{-1.6}\, dy. \tag{39}$$

Under the same setting with Example 1, we obtain the observations on MET $\mathbf{U}_{ob}$ from the simulated data set by $Au = -1$.

Similar to Example 1, after implementing the procedure proposed in Sec. III, we can obtain the learned $(\sigma_L, \alpha_L) = \min_{\sigma \geq 0, \alpha \in (0,2)} G(\sigma, \alpha) = (0.5, 0.6)$. The variation of the error objective function $G(\sigma, \alpha)$ with $(\sigma, \alpha)$ can be seen in Fig. 3(a). Then, we obtain the sparse coefficient $\tilde{\mathbf{q}}$ of the learned drift $\tilde{f}(x)$ with the learned $(\sigma_L, \alpha_L)$ present in table 2, which shows that the learning result is very close to the true value. Meanwhile, we present the comparison between the mean exit time from the learned SDE and the observations in Fig. 3(b), which shows that they are very well in agreement.

**Table 2**: Identified coefficients for the drift term

| Basis | True | Learning |
|---|---|---|
| 1 | 0 | 0 |
| x | 1 | 1.0254 |
| $x^2$ | 0 | 0 |
| $x^3$ | -5 | -5.0400 |
| $x^4$ | 0 | 0 |

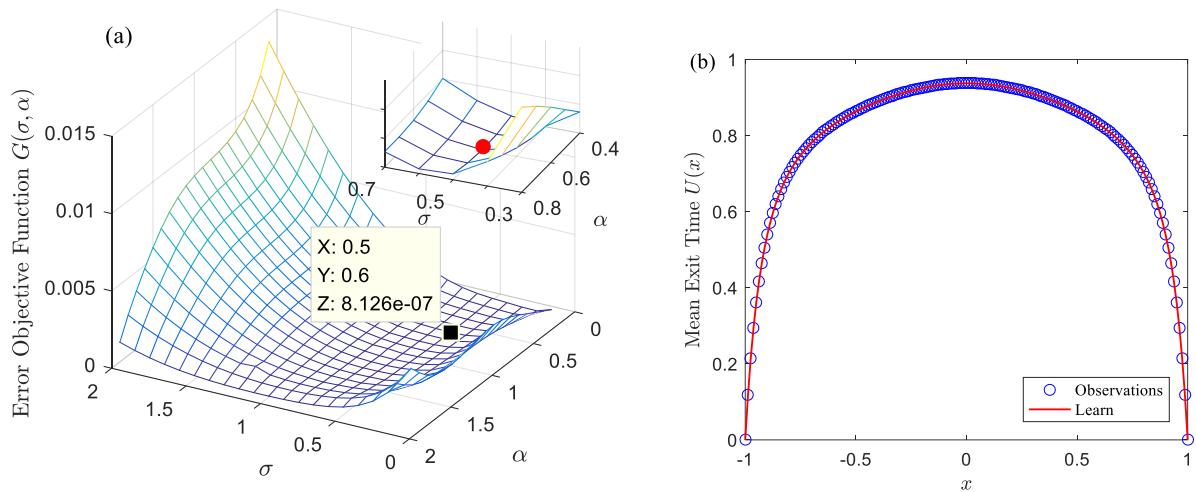

**FIG. 3**. A bi-stable dynamical system with Lévy motion. (a) The error objective function versus $\sigma$ and $\alpha$. (b) Comparation between the mean exit time from the learned SDE and the observations.

## V. DISCUSSION

We have devised a method for extracting governing laws with data only on mean exit time, by solving an inverse problem for a nonlocal partial differential equation and minimizing the error objective function. We present three examples to verify our method with the simulated data from the original systems. Results show that the proposed method apply to not only the dynamical systems driven by Gaussian Brownian motion but also those driven by non-Gaussian Lévy motion. Furthermore, it is still efficient to identify the stochastic dynamical systems with complex rational drift. Currently, most researchers focused on systems driven by Gaussian Brownian motion with observations on system state time series. If it is hard to observe the system state time series or is relatively easy to observe the mean exit time, our method can make a certain contribution to identifying the stochastic differential equations no matter for the dynamical systems driven by either Gaussian Brownian motion or non-Gaussian Lévy motion. With the learned stochastic differential equations, we can further study the stochastic dynamical behaviors of the systems, such as stochastic bifurcation, stochastic resonance and chaotic phenomenon.

We consider the scalar stochastic dynamical equations in the present paper. In fact, our method also applies to the higher dimensional systems in which case the calculating process will be more complex. There also exists a challenge in this method, that is, how to determine the types of Brownian motion and Lévy motion in advance.

## ACKNOWLEDGEMENTS

This study is supported by the National Natural Science Foundation of China (Nos. 11772048, 11832005). The first author warmly acknowledges the financial support of the China Scholarship Council (CSC No. 201906030059) and Excellent Doctoral Dissertation Seedling Fund of Beijing Institute of Technology (BIT) to enable this work, and also wants to thank Dr. Ting Gao and Dr. Xiaoli Chen for helpful discussions.**REFERENCES**

[1] J. Duan, "An Introduction to Stochastic Dynamics," Cambridge University Press, New York (2015).

[2] Y. Zhang, and Y. Jin, "Stochastic dynamics of a piezoelectric energy harvester with correlated colored noises from rotational environment," Nonlinear Dyn. **98** (1), 501-515 (2019).

[3] X. Cheng, H. Wang, X. Wang, J. Duan, and X. Li, "Most probable transition pathways and maximal likely trajectories in a genetic regulatory system," Physica A **531**, 121779 (2019).

[4] S. L. Brunton, J. L. Proctor, and J. N. Kutz, "Discovering governing equations from data by sparse identification of nonlinear dynamical systems," Proc. Natl. Acad. Sci. **113**(15), 3932-3937 (2016).

[5] L. Boninsegna, F. Nüske, and C. Clementi, "Sparse learning of stochastic dynamical equations," J. Chem. Phy. **148**(24), 241723 (2018).

[6] K. Champion, B. Lusch, J. N. Kutz, and S. L. Brunton, "Data-driven discovery of coordinates and governing equations," Proc. Natl. Acad. Sci. **116**(45), 22445-22451 (2019).

[7] A. Mauroy and J. Goncalves, "Linear identification of nonlinear systems: A lifting technique based on the Koopman operator," In 2016 IEEE 55th Conference on Decision and Control (CDC), 6500–6505 (2016).

[8] S. Klus, F. Nüske, S. Peitz, J. H Niemann, C. Clementi, and C. Schütte, Data-driven approximation of the Koopman generator: Model reduction, system identification, and control," Phys. D Nonlinear Phenomena, **406**, 132416 (2020).

[9] N. Takeishi, Y. Kawahara, and T. Yairi, "Subspace dynamic mode decomposition for stochastic Koopman analysis," Phys. Rev. E **96**(3), 033310 (2017).

[10] M. O. Williams, I. G. Kevrekidis, and C.W. Rowley, "A Data Driven Approximation of the Koopman Operator: Extending Dynamic Mode Decomposition," J Nonlinear Sci. **25**(6), 1307-1346 (2015).

[11] Y. Lu, and J. Duan, "Discovering transition phenomena from data of stochastic dynamical systems with Levy noise," arXiv:2002.03280 (2020).

[12] A. J. Chorin, and F. Lu, "Discrete approach to stochastic parametrization and dimension reduction in nonlinear dynamics," Proc. Natl. Acad. Sci. **112**(32), 9804-9809 (2015).

[13] J. Leander, T. Lundh, and M. Jirstrand, "Stochastic differential equations as a tool to regularize the parameter estimation problem for continuous time dynamical systems given discrete time measurements," Math. Biosci. **251**,54-62 (2014).

[14] H. Verdejo, A. Awerkin, W. Kliemann, and C. Becker, "Modelling uncertainties in electrical power systems with stochastic differential equations," Int. J. Elec. Power Energy Syst. **113**, 322-332 (2019).
13

[15] F. Moss, and P. V. E. McClintock, "Noise in nonlinear dynamical systems," UK: Cambridge University Press (2009).

[16] X. Chen, F. Wu, J. Duan, J. Kurths, and X. Li, "Most probable dynamics of a genetic regulatory network under stable Lévy noise," Appl. Math. Comp. **348** 425–436 (2019).

[17] Y. Zheng, F. Yang, J. Duan, X. Sun, L. Fu, and J. Kurths, "The maximum likelihood climate change for global warming under the influence of greenhouse effect and Lévy noise," Chaos **30**, 013132 (2020).

[18] K. Akdim, A. Ez-zetouni, J. Danane, and K. Allali, "Stochastic viral infection model with lytic and nonlytic immune responses driven by Lévy noise," Physica A **549**, 1243671 (2020).

[19] J. G. Lambert, and M. Nezami, "Determination of the Mean Residence Time in the Troposphere by Measurement of the Ratio between the Concentrations of Lead-210 and Polonium-210," Nature **206**, 1343–1344 (1965).

[20] T. R. Jackson, R. Haggerty, S. V. Apte, A. Coleman, and K. J. Drost, "Defining and measuring the mean residence time of lateral surface transient storage zones in small streams," Water Resources Research, **48**(10) (2012).

[21] E. B. Nauman, "Residence time distributions. Handbook of industrial mixing: Science and practice," Wiley Interscience (2004).

[22] O. Levenspiel, "Chemical Reaction Engineering," Wiley, New York (1999).

[23] D. Shin, C. K. Ryu, and S. Choi, "Computational fluid dynamics evaluation of good combustion performance in waste incinerators," J. Air Waste Manage. **48**, 345–351 (1998).

[24] J. A. Novotny, E. C. Parker, S. S. Sruvanshi, G. W. Albin, and L. D. Homer, "Contribution of tissue lipid to long xenon residence times in muscle," J. Appl. Physiol. **74**(5), 2127–34 (1993).

[25] V. Nasserzadeh, J. Swithenbank, D. Lawrence, N. Garrod, and B. Jones, "Measuring gas residence times in large municipal incinerators by means of a pseudo-random binary signal tracer technique," J. Inst. Energ. **68**, 106–120 (1995).

[26] F. Ghirelli, and B. Leckner, "Transport equation for the local residence time of a fluid," Chem. Eng. Sci. **59**, 513–23 (2004).

[27] T. Gao, and J. Duan, "Quantifying model uncertainty in dynamical systems driven by non-Gaussian Lévy stable noise with observations on mean exit time or escape probability," Commun. Nonlinear Sci. Numer. Simulat **39,** 1–6 (2016).

[28] D. Schertzer, M. Larcheveque, J. Duan, V. Yanovsky, and S. Lovejoy, "Fractional Fokker–Planck equation for nonlinear stochastic differential equations driven by non-Gaussian Lévy stable noises," J. Math. Phys. **42**(20), 0–12 (2001).

[29] S. L. Brunton, J. L. Proctor, and J. N. Kutz, "Discovering governing equations from data by sparse identification of nonlinear dynamical systems," Proc. Natl. Acad. Sci. **113**(15), 3932-3937 (2016).

[30] T. Gao, J. Duan, X. Li, and R. Song, "Mean exit time and escape probability for dynamical systems driven by Lévy noise," SIAM J. Sci. Comput. **36**(3), 887–906 (2014).

[31] Q. Liu, and Y. Jia, "Fluctuations-induced switch in the gene transcriptional regulatory system," Phys. Rev. E **70,** 041907 (2004).